\documentclass[a4paper,11pt]{article}
\usepackage[toc,page]{appendix}
\usepackage{amsthm}
\usepackage{amsmath}
\usepackage{latexsym}
\usepackage{amssymb}
\usepackage{verbatim}
\usepackage{setspace}
\usepackage{qtree}
\usepackage{enumitem}
\usepackage{xcolor}
\usepackage{url}
\usepackage{stmaryrd}
\usepackage{authblk}
\usepackage[margin=1.4in]{geometry}

\DeclareMathOperator{\fo}{\mathrm{FO}}
\DeclareMathOperator{\fragment}{\mathcal{L}}

\author{Reijo Jaakkola}
\affil{Tampere University, Finland}

\date{}

\begin{document}

\setlength\abovedisplayskip{3pt}
\setlength\belowdisplayskip{3pt}

\title{An Extension of Trakhtenbrot's Theorem}

\theoremstyle{plain}
\newtheorem{theorem}{Theorem}[section]
\newtheorem{lemma}[theorem]{Lemma}
\newtheorem{corollary}[theorem]{Corollary}
\newtheorem{proposition}[theorem]{Proposition}
\newtheorem{fact}[theorem]{Fact}

\theoremstyle{definition}
\newtheorem{definition}[theorem]{Definition}
\newtheorem{remark}[theorem]{Remark}

\maketitle

\begin{abstract}
\noindent The celebrated Trakhtenbrot's theorem states that the set of finitely valid sentences of first-order logic is not computably enumerable. In this note we will extend this theorem by proving that the finite satisfiability problem of any fragment of first-order logic is \textsc{RE}-complete, as long as it has an effective syntax, it is equi-expressive with first-order logic over finite models and it is effectively closed under conjunction.
\end{abstract}

\section{Introduction}

It is well-known that both the general satisfiability problem and the finite satisfiability problem of first-order logic $\fo$ are undecidable. The general satisfiability problem was proved to be undecidable independently by Church \cite{Church1936ANO} and Turing \cite{Turing1937OnCN}, while the finite satisfiability problem was proved to be undecidable by Trakhtenbrot \cite{trakhtenbrot1950impossibility}, a result which implied the celebrated ``Trakhtenbrot's theorem" which states that the set of finitely valid sentences of $\fo$ is not effectively enumerable. Given these negative results, research efforts have been focused on \emph{fragments} of $\fo$, which are essentially subsets of $\fo$ obtained by imposing further syntactical restrictions on the sentences of $\fo$. The main idea behind this type of research is that there is a fundamental trade-off between expressive power and tractability, and the goal is to find fragments which have both high expressive power and tractable satisfiability problem. For a survey of this research area we direct the reader to \cite{borger97} and to the introduction of \cite{preprintofthis2}.

Given this goal, it is natural ask the following naïve question: is there a fragment of $\fo$ which has the same expressive power as $\fo$ and which furthermore has a decidable satisfiability problem? If we impose no restrictions on what counts as a fragment, then the answer is trivially true: one can, for instance, take all the satisfiable sentences of $\fo$ together with $\bot$ (to make the resulting fragment equi-expressive with $\fo$). However, if we make the very minimal requirement that the fragment should at least have an effective syntax, by which we mean that we can effectively determine whether a given sentence of $\fo$ belongs to this fragment, then one can show quite easily that such a fragment can not have a decidable satisfiability problem. This result follows from the fact that the set of valid sentences of $\fo$ is \textsc{RE}-complete, which allows one to translate $\fo$ effectively into any equi-expressive fragment with an effective syntax.

Thus in the context of general satisfiability problem there is indeed a necessary trade-off between expressive power and tractability. Next we move our attention to the finite satisfiability problem, which has been gaining more and more interest in the recent years due to its applications in computer science. For example, if one views a sentence as a \emph{specification}, then one might be only interested in violations of this specification which are finite. For several fragments these two questions coincide, since they have the finite model property\footnote{Every satisfiable sentence has a finite model.}, but there are also several interesting fragments which do not have this property, such as the two-variable logic with counting \cite{PH05} and the extension of Ackermann fragment with a single unary function \cite{Shelah1977DecidabilityOA}.

Now we can ask an analogous question: is there a fragment of $\fo$ which has the same expressive power as $\fo$ and which furthermore has a decidable finite satisfiability problem? Again, to avoid banal counterexamples, we restrict our attention to fragments that have an effective syntax. By Trakhtenbrot's theorem, the set of finitely valid sentences of $\fo$ is \textsc{coRE}-complete, and so we can not in general translate effectively sentences of $\fo$ into sentences of an arbitrary equi-expressive fragment with an effective syntax. In fact, it is easy to show that in the finite case the undecidability of the fragment does not follow from the fact that it has an effective syntax and it is equi-expressive with $\fo$ over finite models. Indeed, the following set
\[\{\bot\} \cup \{\phi^n \mid \phi \in \fo \text{ and $\phi$ has a model of size $n$}\},\]
where 
\[\phi^n := \underbrace{\phi \land \dots \land \phi}_{n\text{-times}},\]
is a fragment that has an effective syntax, is equi-expressive with $\fo$ over finite models and it has a decidable finite satisfiability problem.

The main purpose of this note is to show that we can nevertheless prove the following.

\begin{theorem}\label{thm:main-theorem}
    If $\mathcal{L}\subseteq \fo$ is a computable set which has the same expressive power as $\fo$ over finite models and is effectively closed under conjunction, then its finite satisfiability problem is \textsc{RE}-complete.
\end{theorem}

The assumptions that we are working with here are still very minimal, since practically all the well-studied decidable fragments of $\fo$ are effectively closed under conjunction. Furthermore, one could argue that at least from the point of view of applications, a decidable fragment is not very useful if it is not effectively closed under conjunction, since then one can not write down constraints in a \emph{modular} fashion.

\section{Preliminaries}

We fix some reasonable enumeration $(M_x)_{x\in \{0,1\}^*}$ of all Turing machines. In particular, the binary string $x$ encodes the Turing machine $M_x$. We also fix some computable set $\mathcal{V}$ of relation symbols, which will serve us as a background vocabulary. We will use $\fo$ to denote the set of all sentences of first-order logic over the vocabulary $\mathcal{V}$.

A computable subset $\fragment \subseteq \fo$ is a \emph{fragment} of $\fo$. Most of them are closed under conjunction syntactically, by which we mean that if $\phi$ and $\psi$ are sentences of a fragment then so is $\phi \land \psi$, but there are also those fragments which are only closed under conjunction with respect to expressive power (for instance some of the prefix fragments). This motivates the following definition: we say that a fragment $\fragment$ of $\fo$ is \emph{effectively closed under conjunction}, if there exists a computable mapping $f:\fragment \times \fragment \to \fragment$ so that $f(\phi,\psi)$ is equivalent with $\phi \land \psi$. 

\emph{Word} is a finite structure $\mathfrak{A}$ over a vocabulary consisting of unary relation symbols and a single binary relation symbol $<$ which satisfies the following two requirements.
\begin{enumerate}
    \item $<^\mathfrak{A}$ is a linear ordering of $A$.
    \item For every $a\in A$ there exists precisely one unary relation symbol $P$ in the underlying vocabulary for which $a\in P^\mathfrak{A}$.
\end{enumerate}
The following well-known fact is proved in \cite{stockmeyer1974complexity}.

\begin{proposition}\label{prop:fo-over-words}
    The satisfiability problem of $\fo$ over words is decidable.
\end{proposition}

\section{Proof of the main result}

Let $\fragment$ be a fragment of $\fo$ that has the same expressive power as $\fo$ over finite models and is effectively closed under conjunction. Our goal is to show that we can effectively reduce halting problem to the finite satisfiability problem of $\fragment$. 

The chief technical obstacle that we need to bypass is that over finite models we can not in general translate effectively $\fo$-sentences to sentences of $\fragment$. However, by Proposition \ref{prop:fo-over-words}, we know that we can effectively determine whether two sentences $\fo$ are equivalent over words, and hence any such $\fo$-sentence $\phi$ can be effectively translated to a sentence $\phi' \in \fragment$ which is equivalent with $\phi$ over words. Our general strategy is now that we will use these sentences to encode inputs to a universal Turing machine, which in turn will be encoded with a single \emph{fixed} sentence of $\fragment$.

Now we will proceed with the formal proof. Let $U$ be a Turing machine, which when given as an input a binary string $x$, simulates the run of $M_x$ on empty input. Thus $U$ halts on $x$ if and only if $M_x$ halts on empty input. We will assume that the vocabulary of $U$ is $\{0,1\}$ and that $U$ has a single one-way infinite memory tape. We will next describe sentences $\phi_U$ and $\phi_x$ which have the following properties.

\begin{enumerate}
    \item $\phi_U \land \phi_x$ is finitely satisfiable iff $U$ halts on input $x$.
    \item For every model $\mathfrak{A}$ of $\phi_U$ the restriction of $\mathfrak{A}$ to the vocabulary of $\phi_x$ is a word.
\end{enumerate}

\noindent We start with the sentence $\phi_x$. Consider the vocabulary $\{E,Z,O,<\}$, where $E,Z,O$ are unary while $<$ is binary. Intuitively $Z$ and $O$ are used to indicate that an element is labelled with zero and one respectively, while $E$ is used to indicate that the element is labelled with a ``blank" symbol, i.e., it correspondence to an empty square on the tape of $U$. $\phi_x$ will be the following sentence:
\[\exists x_1 \dots \exists x_n [\mathrm{first}_<(x_1) \land \bigwedge_{1\leq i < n} \mathrm{succ}(x_i,x_{i+1})\] 
\[\land \bigwedge_{x_i = 0} Z(x_i) \land \bigwedge_{x_i = 1} O(x_i) \land \forall y (x_n < y \to E(y))].\]
Here $\mathrm{first}_<(z)$ describes the fact that $z$ is the smallest element in $<$, while $\mathrm{succ}(z,w)$ describes the fact that $w$ is the immediate successor of $z$ in $<$. Thus, over words, the sentence $\phi_x$ expresses the fact that the underlying word consists of $x$ which is followed by a sequence of ``blank" symbols.

Next we will consider the sentence $\phi_U$. Before describing it, we need to discuss how we are going to formalize the computations of $U$. In brief, we will describe the computation of $U$ as a finite grid, where the $i$:th row describes a sufficiently large prefix of the memory tape of $U$ at stage $i$. In addition to $\tau_0$, the vocabulary of $\phi_U$ will contain in addition at least three binary relation symbols $P_{zero},P_{one},P_{empty}$, which can be used to describe the content of the memory tape of $U$; and a binary relation symbol $<'$ which will be used to describe a second linear ordering on the models domain. Roughly speaking, in the finite grid the element at $(i,j)$ correspondence to a pair of elements in the domain where the first element is the $i$:th element in $<$ while the second element is the $j$:th element in $<'$.

Given this intuition, it is mostly routine to write down $\phi_U$. The only non-standard part of $\phi_U$ is the manner in which we encode the input received by $U$, since we want $\phi_U$ to be independent of $x$. The following sentence
\[\exists x [\mathrm{first}_{<'}(x) \land \forall y ((Z(y) \to P_{zero}(x,y)) \land (O(y) \to P_{one}(x,y)) \land (E(y) \to P_{empty}(x,y)))],\]
where $\mathrm{first}_{<'}(x)$ is a formula specifying that $x$ is the first element in the linear order $<'$, will take care of this, since together with the sentence $\phi_x$ it encodes the fact that the bottom row in the finite grid -- essentially the input tape of $U$ -- starts with the binary word $x$ followed by a sequence of empty squares.

To complete the proof, we will describe how a sentence of $\fragment$ equivalent with $\phi_U \land \phi_x$ can be obtained effectively from the input string $x$. First, since the sentence $\phi_U$ does not depend on the word $x$, we can hardcode an equivalent sentence $\phi_U' \in \fragment$ into our procedure. Secondly, since the satisfiability problem of $\fo$ over words is decidable, we can effectively find a sentence of $\fragment$ which is equivalent with the sentence $\phi_x$ over words; let $\phi_x' \in \fragment$ denote this sentence. Since $\mathcal{L}$ is effectively closed under conjunction, we can effectively compute a third sentence $\chi_x$ which is equivalent with $\phi_U' \land \phi_x'$.

\section{Conclusions}

In this note we have proved that there exists no computable subset of the set of sentences of $\fo$ which is equi-expressive with $\fo$ over finite models, is effectively closed under conjunction and furthermore has a decidable finite satisfiability problem. This result extends the celebrated Trakhtenbrot's theorem, since it essentially shows that the undecidability of the finite satisfiability problem of $\fo$ can not be avoided by simply fixing the syntax of $\fo$ in an appropriate way, if we require that the resulting syntax should satisfy some very minimal requirements. More generally speaking, our result shows that the the tight connection between expressive power and decidability, which in the case of general models follows from the fact that valid sentences of $\fo$ can be enumerated by a Turing machine, extends also to the case of finite models.

\section*{Acknowledgement}

This note is based on ideas that I developed while I was writing my master's thesis. I would like to thank my advisors Antti Kuusisto and Lauri Hella for their useful and encouraging comments on these ideas.

\bibliographystyle{plainurl}
\bibliography{arxiv}

\begin{thebibliography}{1}

\bibitem{borger97}
Egon B{\"{o}}rger, Erich Gr{\"{a}}del, and Yuri Gurevich.
\newblock {\em The Classical Decision Problem}.
\newblock Perspectives in Mathematical Logic. Springer, 1997.

\bibitem{Church1936ANO}
Alonzo Church.
\newblock A note on the entscheidungsproblem.
\newblock {\em J. Symb. Log.}, 1:40--41, 1936.

\bibitem{preprintofthis2}
Reijo Jaakkola and Antti Kuusisto.
\newblock Algebraic classifications for fragments of first-order logic and
  beyond.
\newblock {\em arXiv Preprint}, arXiv:2005.01184v2, 2021.

\bibitem{PH05}
Ian Pratt-Hartmann.
\newblock Complexity of the two-variable fragment with counting quantifiers.
\newblock {\em Journal of Logic, Language and Information}, 14(3):369--395,
  2005.

\bibitem{Shelah1977DecidabilityOA}
Saharon Shelah.
\newblock Decidability of a portion of the predicate calculus.
\newblock {\em Israel Journal of Mathematics}, 28:32--44, 1977.

\bibitem{stockmeyer1974complexity}
Larry Stockmeyer.
\newblock {\em The Complexity of Decision Problems in Automata Theory and
  Logic}.
\newblock Massachusetts Institute of Technology, Project MAC, 1974.

\bibitem{trakhtenbrot1950impossibility}
Boris Trakhtenbrot.
\newblock Impossibility of an algorithm for the decision problem on finite
  classes (in russian).
\newblock {\em Doklady Akademii Nauk SSSR}, 70:569--572, 1950.

\bibitem{Turing1937OnCN}
Alan Turing.
\newblock On computable numbers, with an application to the
  entscheidungsproblem.
\newblock {\em Proceedings of The London Mathematical Society}, 41:230--265,
  1937.

\end{thebibliography}

\end{document}